\documentclass[journal]{IEEEtran}
\usepackage{graphicx}
\usepackage{epstopdf}
\usepackage[latin1]{inputenc}
\usepackage{amsmath,amssymb,amscd,latexsym,dsfont}
\usepackage[english]{babel}
\usepackage{cite}
\usepackage{epstopdf}
\usepackage{graphicx}
\usepackage{algpseudocode}
\usepackage{algorithm}
\usepackage{algorithmicx}
\usepackage{float}
\usepackage{multicol}
\usepackage{psfrag}
\usepackage{comment,psfrag,subfigure,enumerate}
\usepackage{color}
\usepackage{amsthm}

\ifCLASSINFOpdf
\else
\fi
\begin{document}
%
\title{An Error-limited NOMA-HARQ Approach using Short Packets}

\author{
    \IEEEauthorblockN{Behrooz Makki, ~\IEEEmembership{Senior Member,~IEEE}, Tommy Svensson,~\IEEEmembership{Senior Member,~IEEE}, Michele Zorzi,~\IEEEmembership{Fellow,~IEEE}}
    \thanks{Behrooz Makki is with Ericsson Research, Ericsson, 417 56 Gothenburg, Sweden, Email: behrooz.makki@ericsson.com.}
    \thanks{Tommy Svensson is with the Electrical Engineering Department, Chalmers University of Technology, 412 96 Gothenburg, Sweden, e-mail:
tommy.svensson@chalmers.se.}
\thanks{Michele Zorzi is with the Department of Information Engineering, University of Padova, 35122 Padova, Italy, e-mail: zorzi@dei.unipd.it.}
    
}
\maketitle

\begin{abstract}
In this paper, we propose and analyze an error-constrained data transmission approach for uplink communication {using} non-orthogonal multiple access (NOMA). The results are presented for the cases with hybrid automatic repeat request (HARQ) protocols where, depending on the message decoding status in the previous rounds, the decoding scheme of the receiver as well as the data rates/transmission powers of the transmitters are adapted correspondingly. We perform finite block-length analysis of the proposed scheme and study the effect of the codewords length on the performance of NOMA-HARQ protocols. Particularly, considering repetition time diversity HARQ and codewords of finite length, we determine the expected error-limited achievable rate and the transmit power improvement of our proposed scheme, compared to the conventional NOMA-HARQ protocols. As we show, smart adaptation of the decoding scheme improves the spectral/energy efficiency of NOMA-HARQ protocols considerably with no additional receiver complexity.
\end{abstract}



%
\IEEEpeerreviewmaketitle
\vspace{-4mm}
\section{Introduction}
The design of multiple access (MA) schemes is of interest in wireless networks. Here, the goal is to provide user equipments (UEs) with radio resources in a spectrum-, cost- and complexity-efficient manner. In 1G-3G, frequency division multiple access (FDMA), TDMA (T: time) and CDMA (C: code) schemes have been introduced, respectively. Then, long-term evolution (LTE) and LTE-Advanced developed orthogonal frequency division multiple access (OFDMA) and single-carrier (SC)-FDMA as OMA (O: orthogonal) schemes. Such orthogonal designs have the benefit that there is no mutual interference among UEs\footnote{{In practice, there is some limited controlled residual interference, e.g., from adjacent channel leakage and, indeed, from inter-cell interference.}}, leading to high system performance with simple receivers.

Recently, non-orthogonal multiple access (NOMA) has received considerable attention as a promising MA technique {for certain scenarios, e.g., asynchronous access in massive machine type communications (see \cite{8972353} for a survey on the potentials and challenges of NOMA)}. With NOMA, multiple UEs are paired and share the same radio resources in time, frequency and/or code. Various fundamental results have been presented to determine the ultimate performance of NOMA in both downlink (DL) \cite{7812683,8558537,poornoma,7542118,7433470}
and uplink (UL) \cite{7433470,7542118,7390209,7913656}.
These results have been presented for the cases with asymptotically long codewords, where the achievable rates are given by the Shannon's capacity formula. However, finite block length analysis of NOMA has been rarely considered in, e.g., \cite{8345745,8644329,8292490,arxivUPnoma}, where the results are presented mainly for DL  {scenarios} \cite{8345745,8644329,8292490}.

In 3GPP, NOMA has been considered for different applications. For instance, NOMA has been introduced as an extension {of} network-assisted interference cancellation and suppression (NAICS) for inter-cell interference (ICI) mitigation in LTE Release 12 \cite{refnoma1}, as well as a study-item of LTE Release 13, under the name of DL multi-user superposition transmission (DMST) \cite{refnoma2}. Also, recently a study-item on NOMA has been considered in Release 15, in which the performance of various NOMA techniques and their practical implementation challenges have been investigated \cite{8972353,3gppfinalnoma}. 

Different techniques have been proposed to incorporate  {typical} data transmission methods such as hybrid automatic repeat request (HARQ) to the cases using NOMA \cite{7501524,8408492,8335325,8911437,8963920,8884744}, to develop low-complexity UE pairing schemes \cite{7557079,8016604,7511620}, and to reduce the receiver complexity \cite{8423459,8630078,7752784}. 
As shown in these works, with proper parameter settings, NOMA has the potential to outperform {existing} OMA techniques at the cost of {advanced} receiver, UE pairing and coordination complexity. For these reasons, NOMA has been suggested as a candidate for data transmission in dense networks with a large number of UEs requesting {access, where} there are not enough orthogonal resources to serve them in an OMA-based fashion.

Due to the channel state information (CSI) acquisition and UE pairing overhead, NOMA is of most interest in fairly static channels with no frequency hopping where the channels remain constant for a number of packet transmissions. In such cases, the network suffers from poor diversity. Moreover, NOMA is faced with error propagation problems where, if the receiver fails to decode a signal, its interference affects the decoding probability of all remaining signals which should be decoded sequentially. For these reasons, there may be a high probability {of} requiring HARQ-based retransmissions. On the other hand, as we show in this paper, the unique properties of NOMA give {a} chance to improve the performance of paired UEs during HARQ-based data transmissions, if the transmission parameters and the decoding schemes are adapted smartly. This is the motivation for our work, in which we develop an efficient NOMA-HARQ protocol for UL communications.


In this paper, we develop and analyze an error-constrained UL NOMA scheme using HARQ. The objective is to improve the error-limited rates of one of the paired UEs, or equivalently, improve its energy efficiency, with no additional complexity at the receiver and no penalty for the other paired UEs. In our proposed setup, depending on the message decoding status of the paired UEs, the receiver adapts its message decoding scheme. Moreover, the receiver informs each UE about the message decoding status of all paired UEs and the adapted message decoding scheme, such that each UE updates its transmission parameters, e.g., rate {and} power, correspondingly.

We use the fundamental results on the achievable rates of finite block-length codes \cite{5452208,letterzorzikhodemun,7037211,7572902,8964482,7463506} to analyze the system performance in the cases with short packets. With codewords of finite length, we study the error-limited expected rate improvement and the transmission power reduction of the proposed scheme, compared to conventional NOMA-HARQ protocols. Here, the results are presented for the cases with repetition time diversity (RTD) HARQ \cite{throughputdef,6566132,6164088,7438859,5771499}, while the same approach is applicable for other HARQ protocols as well. Finally, we study the effect of different parameters such as the codeword length and the UEs' error constraints on the system performance.

Our analytical and simulation results show that
  with different codeword lengths and a broad range of error constraints, the expected achievable rates of both the standard and {the} proposed NOMA-HARQ protocols scale with the error probability, in the log domain, almost linearly.
  Also, the spectral and energy efficiency of the error-limited NOMA-HARQ based UEs are sensitive to the error constraints and the codewords length, in the cases with short packets. However, the sensitivity to the error constraint/codeword length decreases as the codeword length increases. 
  Finally, adapting the decoding scheme in different transmission rounds of HARQ gives {a} chance to improve the spectral and energy efficiency of NOMA-HARQ protocols significantly. For instance, consider codewords of length 1000 channel uses (cu) and rate 1 npcu (np: nats per), Rayleigh-fading channels with typical parameter settings for two paired cell-edge and cell-center UEs, and error constraint $10^{-3}$ for both UEs. Then, compared to standard NOMA-HARQ, our proposed scheme reduces the required signal-to-noise ratio (SNR) of the cell-center UE in the retransmissions by {$9$} dB.     

\begin{figure*}
\vspace{-3mm}
\centering
  \includegraphics[width=1.96\columnwidth]{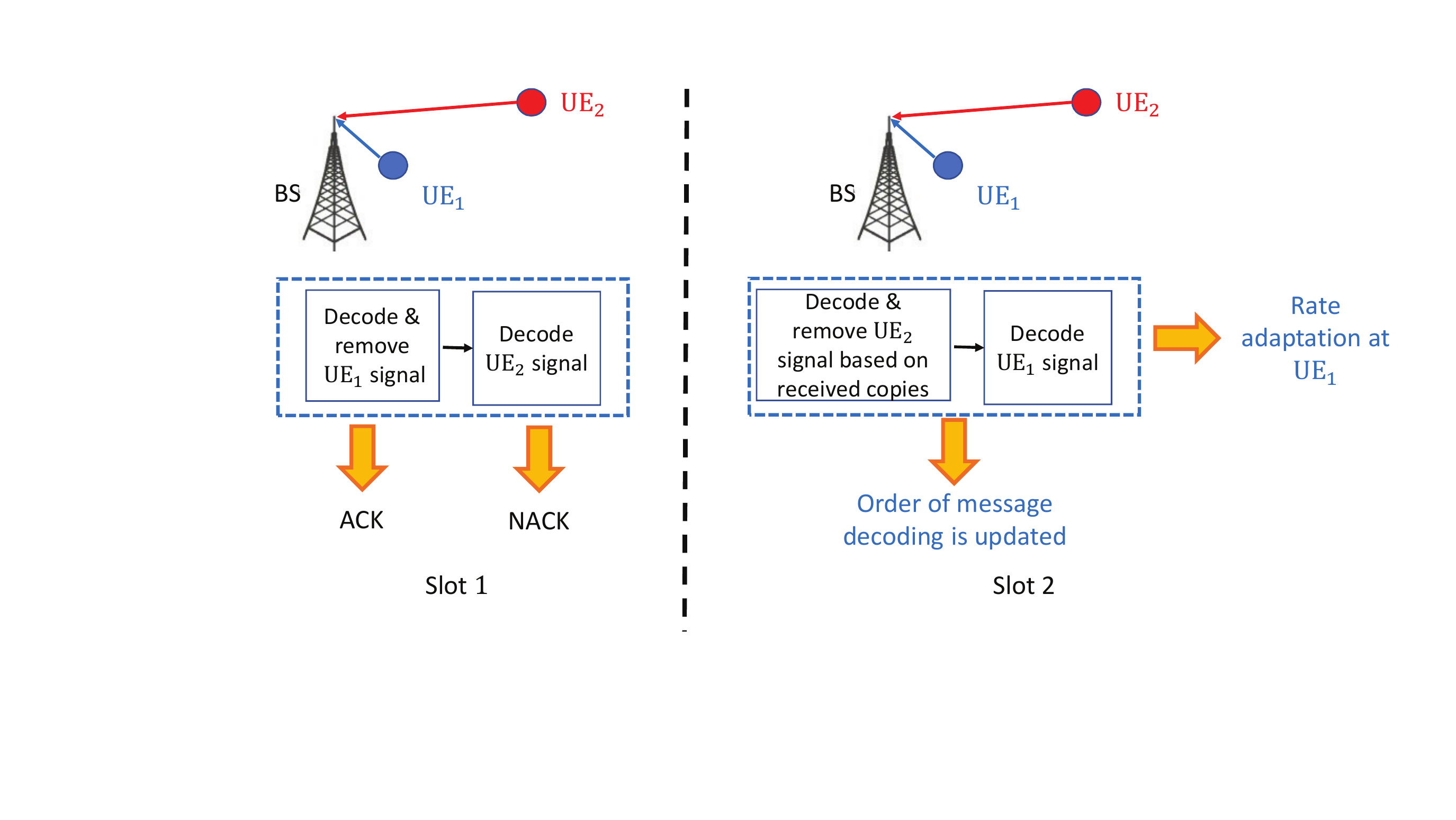}\\\vspace{-20mm}
\caption{Schematic of the proposed setup. If the message of the cell-center UE is correctly decoded in Slot 1 while the BS fails to decode the message of the cell-edge UE, in time Slot 2 the BS changes the order of the message decoding in the SIC-based receiver. While with HARQ the changed order of message decoding still guarantees the error probability constraint of the cell-edge UE, it gives the cell-center UE {a} chance to improve its performance.}
\vspace{-0mm}
\end{figure*}

\section{System model}

Consider the simplest case of {a} UL scenario where two cell-edge and cell-center UEs, i.e., two UEs with {widely different} channel qualities, {want} to connect to a base station (BS). {As} we explain in Section III, the same approach can be applied to the cases with different numbers of paired UEs. Let us denote the cell-center and the cell-edge UEs by $\text{UE}_1$ and $\text{UE}_2$, respectively. With NOMA, the UEs share the same spectrum and time resource to send their messages simultaneously. Thus, the signal received by the BS in time slot $t$ is {proportional to}
\begin{align}\label{eq:eqsystemmodel}
Y(t)=\sqrt{P_1}H_1(t)X_1(t)+\sqrt{P_2}H_2(t)X_2(t)+Z(t).
\end{align}
Here, $X_i(t)$ is the unit-variance transmission signal of $\text{UE}_i, i=1,2,$ $H_i$ represents the fading coefficient in the $\text{UE}_i$-BS link and $Z \sim \mathcal{CN}(0,1)$ is the independent and identically distributed (IID) complex Gaussian noise added at the BS. Also, $P_i$ denotes the transmit power of $\text{UE}_i, i=1,2.$ Finally, using successive interference cancellation (SIC)-based receiver, the BS decodes the messages sequentially.




The results are presented for the cases with quasi-static channels where the channel coefficients remain constant during each transmission round. This is an appropriate model {for} scenarios with slow-moving or stationary users, e.g., \cite{6164088,7438859,5771499}. However, as we explain in Section III, the same approach is also applicable for the cases with fast-fading conditions. Let us define $G_1\doteq |H_1|^2$ and $G_2\doteq |H_2|^2$ which are referred to as the channel gains in the following. Also, we denote the probability density function (PDF) and the cumulative distribution function (CDF) of the random variables $G_i, i=1,2,$ by $f_{G_i}(\cdot)$ and $F_{G_i}(\cdot)$, respectively. We consider Rayleigh-fading conditions with $f_{G_i}(x)=\lambda_ie^{-\lambda_ix}, i=1,2,$ where $\lambda_i, i=1,2,$ depends on the long-term channel quality/path loss. Also, we assume the channel coefficients to be known by the transmitters and the BS which is an acceptable assumption in quasi-static conditions. Finally, the results are presented for the cases with RTD HARQ where the same signal (with possibly different power) is sent in each retransmission round, and the receiver performs maximum ratio combining of all received copies of the signal.

\section{Proposed Scheme}

Our goal is to improve the achievable rate of the paired UEs in error-limited conditions. Considering the transmission setup of Fig. 1 in Slot 1, the error probability for the data transmission of $\text{UE}_i, i=1,2,$ is given by
\begin{align}\label{eq:eqE2EDF}
\left\{\begin{matrix}
\Theta_1=\mathcal{F}\left(r_1,\frac{G_1P_1}{1+G_2P_2} \right ), \,\,\,\,\,\,\,\,\,\,\,\,\,\,\,\,\,\,\,\,\,\,\,\,\,\,\,\,\,\,\,\,\,\,\,\,\,\,\,\,\,\,\,\,\,\,\,\,\,\,\,\,\,\,\,\,\,\,\,\,\,\,\,\,\,\,\,\,\,\,\,\text{(i)}\\
\Theta_2=(1-\Theta_1)\mathcal{F}\left(r_2,G_2P_2 \right )+\Theta_1\mathcal{F}\left(r_1,\frac{G_2P_2}{1+G_1P_1} \right ), \,\,\,\,\text{(ii)}
\end{matrix}\right.
\end{align}
where $r_i$ represents the data rate of $\text{UE}_i,i=1,2$, and $\mathcal{F}(r_i,U_i )$ is a function of the considered coding scheme, data rate and the useful signal power, e.g., signal-to-interference-plus-noise ratio (SINR) or the SNR, depending on the order and the status of message decoding of the paired signals. Moreover, (2.i) comes from the fact that, using the SIC receiver, the BS decodes the message of $\text{UE}_1$ in the presence of interference. Also, (2.ii) is based on the fact that, as opposed to the cases with infinitely long codewords, with short packets the BS can still try decoding the signal of $\text{UE}_2$ even if it fails to decode and remove the signal of $\text{UE}_1.$

Note that the error function $\mathcal{F}(r_i,U_i )$ is an increasing function of $r_i$ and a decreasing function of $U_i.$ In this work, we consider the fundamental results of \cite{5452208} where, considering the ultimate performance for the codewords of moderate length,  $\mathcal{F}(r_i,U_i )$ is given by
\begin{align}\label{eq:eqerrorpoly}
\mathcal{F}(r_i,U_i )=Q\left(\frac{\sqrt{L}\left(\log(1+U_i)-r_i\right)}{\sqrt{1-\frac{1}{(1+U_i)^2}}}\right).
\end{align}
Here, $L$ is the codewords length, $Q(x)=\frac{1}{\sqrt{2\pi}}\int_x^\infty{e^{-\frac{t^2}{2}}\text{d}t}$ denotes the $Q$ function, and we have $r_i=\frac{K_i}{L}, i=1,2,$ where $K_i$ represents the number of information nats of the codewords considered by $\text{UE}_i$.

NOMA-based transmission in quasi-static dense networks has advantages and disadvantages. As a disadvantage, the network may require multiple message retransmissions, due to the lower average SINR compared to typical OMA systems and also the static behavior of the channel coefficients, which affects the performance of HARQ protocols significantly. Particularly, as the relative performance gain of NOMA increases when two UEs with different channel qualities are paired, there is a high probability that the cell-edge UE may need retransmissions. On the other hand, along with low CSI/UE pairing overhead, one of the main advantages of NOMA-based transmission in stationary networks is that the system performance can be predicted with high accuracy, which makes it possible to apply smart data transmission techniques. We exploit this point to improve the achievable rate/energy efficiency of the cell-center UE as follows.

Figure 1 shows the schematic of our proposed NOMA-HARQ setup. To simplify the discussions, we present the setup for the simplest case with two (re)transmission rounds, two paired UEs and UL transmission. However, as we explain in the following, the same approach is also applicable for the cases with different numbers of retransmissions/HARQ protocols and DL transmission. Let us denote the maximum error probability accepted for the data transmission of $\text{UE}_i$ by $\theta_i$. That is, the data transmission is constrained to $\Theta_i\le \theta_i,i=1,2.$ Moreover, we concentrate on the case where, while the BS decodes the message of the cell-center UE, i.e., $\text{UE}_1$, correctly, it can not decode the message of the cell-edge UE, i.e., $\text{UE}_2$, in the first transmission round. Discussions on the system performance with other possible cases of message decoding conditions are presented in the sequel. In this way, {we propose that the} data transmission scheme follows the following procedure (see Fig. 1):

\begin{itemize}
  \item At the BS: In Slot 1, the BS uses the CSI and the UEs maximum error constraints $\theta_i, i=1,2,$ to determine the appropriate data rate of the UEs and the order of message decoding according to
\begin{align}\label{eq:eqrate1}
&\left\{\begin{matrix}
r_1=\arg_x\left\{\mathcal{F}\left(x,\frac{P_1G_1}{1+P_2G_2} \right )=\theta_1\right\}\,\,\,\,\,\,\,\,\,\,\,\,\,\,\,\,\,\,\,\,\,\,\,\,\,\,\,\,\,\,\,\,\,\,\,\,\,\,\,\,\,\,\,\,\,\,\,\,\,\,\,\,\,\,\,\,\,\,\,\,\,\\
r_2=\arg_x\bigg\{(1-\theta_1)\mathcal{F}\left(x,P_2G_2 \right )\,\,\,\,\,\,\,\,\,\,\,\,\,\,\,\,\,\,\,\,\,\,\,\,\,\,\,\,\,\,\,\,\,\,\,\,\,\,\,\,\,\,\,\,\,\,\,\,\,\,\,\,\,\,\,\,\,\,\,\,\,\,\,\\+\theta_1 \mathcal{F}\left(x,\frac{P_2G_2}{1+P_1G_1} \right )=\theta_2\bigg\}
\end{matrix}\right.
\nonumber\\&
\mathop  = \limits^{(a)}
\left\{\begin{matrix}
r_1=\arg_x\left\{Q\left(\frac{\sqrt{L}\left(\log(1+\frac{P_1G_1}{1+P_2G_2})-x\right)}{\sqrt{1-\frac{1}{(1+\frac{P_1G_1}{1+P_2G_2})^2}}}\right)=\theta_1\right\}\\
r_2=\arg_x\Bigg\{(1-\theta_1)Q\left(\frac{\sqrt{L}\left(\log(1+P_2G_2)-x\right)}{\sqrt{1-\frac{1}{(1+P_2G_2)^2}}}\right)\\+\theta_1Q\left(\frac{\sqrt{L}\left(\log(1+\frac{P_2G_2}{1+P_1G_1})-x\right)}{\sqrt{1-\frac{1}{(1+\frac{P_2G_2}{1+P_1G_1})^2}}}\right)=\theta_2\Bigg\},
\end{matrix}\right.
\end{align}
where $(a)$ comes from (\ref{eq:eqerrorpoly}).
{After} finding $r_i$'s,  the BS sends feedback to the UEs to inform each UE about the channel quality of both UEs, the considered order of message decoding, i.e., if a UE is considered as a cell-edge or cell-center UE, as well as the maximum rate such that the UEs error probability constraints $\Theta_i\le\theta_i,i=1,2,$ are satisfied.
  \item At the UEs: Based on the received information, the UEs adapt their transmission rates and send the information messages in the same spectrum resources as given in (\ref{eq:eqsystemmodel}).
  \item At the BS: Assume that the message of $\text{UE}_1$ (resp. $\text{UE}_2$) is correctly decoded (resp. not decoded) at the end of Slot 1.
   Thus, in time Slot 2, $\text{UE}_1$ will send a new message while $\text{UE}_2$ needs to retransmit the previous message, and the BS tries to decode the message of $\text{UE}_2$ based on the two copies of its received signals. In this case, we propose that the BS changes the order of message decoding in the SIC-based receiver. This is because, considering the fact that in Slot 1 the rate of $\text{UE}_2$ was set such that its average error constraint is satisfied, adding the second copy in Slot 2 will indeed guarantee its error constraint \emph{even if the second copy of the signal is affected by interference}\footnote{In fact, the retransmission of the failed signal increases the successful message decoding probability for $\text{UE}_2$, because the two received copies of the signal are combined during the decoding process.}. Therefore, in Slot 2, we can first decode the message of $\text{UE}_2$, by, e.g., maximum ratio {combining (MRC)} its two interference-free and interference-affected signals {received} in Slots 1 and 2, respectively. Changing the order of message decoding, on the other hand, gives the cell-center UE, i.e., $\text{UE}_1$, {a} chance to send the new signal with a higher rate. Particularly, following the same arguments as in (\ref{eq:eqrate1}), the data rate of $\text{UE}_1$ in Slot 2 increases to
      \begin{align}\label{eq:eqrate2}
      &\tilde r_1=
      \nonumber\\&\arg_x\left\{\left(1-\tilde \theta_2\right)\mathcal{F}\left(x,P_1G_1 \right )+\tilde\theta_2\mathcal{F}\left(x,\frac{P_1G_1}{1+P_2G_2} \right )=\theta_1\right\}
      \nonumber\\&
      \mathop  \simeq \limits^{(b)}\arg_x\left\{\mathcal{F}\left(x,P_1G_1 \right )=\theta_1\right\}
      \nonumber\\&
      =\arg_x\left\{Q\left(\frac{\sqrt{L}\left(\log(1+P_1G_1)-x\right)}{\sqrt{1-\frac{1}{(1+P_1G_1)^2}}}\right)=\theta_1\right\},
      \end{align}
where $\tilde \theta_2$ denotes the probability of unsuccessful decoding for the message of $\text{UE}_2$ while combining the two copies of its signal. Then,  $(b)$ is based on the fact that, with the considered rate allocation of the cell-edge UE and the combination of its two received signals, with high probability the signal of the cell-edge UE is correctly decoded at the end of Slot 2, i.e., $\tilde \theta_2$ is small. Finally, note that we use this approximation only for simplifying the discussions. However, the same approach is applicable for the cases with no approximation in (\ref{eq:eqrate2}).

In this way, at the beginning of Slot 2, the BS sends acknowledgement (ACK) and negative-acknowledgement (NACK) to $\text{UE}_1$ and $\text{UE}_2$, respectively, to inform them about their message decoding status. Also, it updates its message decoding order and informs $\text{UE}_1$ about its new maximum achievable rate $\tilde r_1$ found by (\ref{eq:eqrate2}).
\item At the UEs: Receiving NACK and considering, e.g., RTD HARQ protocol, $\text{UE}_2$ retransmits the same signal in Slot 2. The cell-center $\text{UE}_1$, however, adapts the data rate based on (\ref{eq:eqrate2}) and sends a new signal in the spectrum resources shared with $\text{UE}_2.$
\end{itemize}
In this way, our proposed scheme makes it possible to increase the achievable rate of the cell-center UE during the retransmission rounds of the cell-edge UE. This is particularly because, using (\ref{eq:eqrate1}), the achievable rate of the cell-center UE in Slot 1 is obtained as
\begin{align}\label{eq:achiverate1}
r_1=\log\left(1+\frac{P_1G_1}{1+P_2G_2}\right)-\frac{Q^{-1}(\theta_1)}{\sqrt{L}}\sqrt{1-\frac{1}{\left(1+\frac{P_1G_1}{1+P_2G_2}\right)^2}},
\end{align}
where $Q^{-1}(\cdot)$ represents the inverse $Q$ function. Then, considering Rayleigh fading conditions with $f_{G_i}(x)=\lambda_i e^{-\lambda_ix},i=1,2,$ the CDF of the SINR term $U_1=\frac{P_1G_1}{1+P_2G_2}$ is given by
\begin{align}\label{eq:sinrcdf}
F_{U_1}(x)&=\int_0^\infty{f_{G_2}(y)\Pr\left(G_1\le\frac{x}{P_1}\left(1+P_2y\right)\right)\text{d}y}\nonumber\\&
=\int_{0}^{\infty}{\lambda_2e^{-\lambda_2y}\left(1-e^{-\frac{\lambda_1x}{P_1}(1+P_2y)}\right)\text{d}y}\nonumber\\&
=1-\frac{e^{-\frac{\lambda_1}{P_1}x}}{1+\frac{\lambda_1P_2}{\lambda_2P_1}x},
\end{align}
and, denoting the expectation operator by $E\{\cdot\}$, the expected achievable rate of $\text{UE}_1$ in Slot 1 is given by
\begin{align}\label{eq:eqexpectedachiverate1}
E\{ r_1\}&=\int_0^\infty{f_{U_1}(x)\bigg(\log(1+x)-\frac{Q^{-1}(\theta_1)}{\sqrt{L}}\sqrt{1-\frac{1}{\left(1+x\right)^2}}\bigg)\text{d}x}
\nonumber\\&
\mathop  \simeq \limits^{(c)}
\int_0^\infty{f_{U_1}(x)\bigg(\log(1+x)-\frac{Q^{-1}(\theta_1)}{\sqrt{L}}\frac{x}{1+x}\bigg)\text{d}x}
\nonumber\\&
\mathop  = \limits^{(d)}
\int_0^\infty{\frac{1-F_{U_1}(x)}{1+x}\text{d}x}-\frac{Q^{-1}(\theta_1)}{\sqrt{L}}\int_0^\infty{\frac{1-F_{U_1}(x)}{(1+x)^2}\text{d}x}\nonumber\\&
=\frac{1}{\frac{\lambda_1P_2}{\lambda_2P_1}-1}\bigg(e^{\frac{\lambda_2}{P_2}}\text{Ei}\left(-\frac{\lambda_2}{P_2}\right)-e^{\frac{\lambda_1}{P_1}}\text{Ei}\left(-\frac{\lambda_1}{P_1}\right)\bigg)
\nonumber\\&
-\frac{Q^{-1}(\theta_1)}{\sqrt{L}\left(\frac{\lambda_1P_2}{\lambda_2P_1}-1\right)^2}\times
\nonumber\\&\Bigg(e^{\frac{\lambda_1}{P_1}}\left(\frac{\lambda_1}{P_1}\left(\frac{\lambda_1P_2}{\lambda_2P_1}-1\right)+\frac{\lambda_1P_2}{\lambda_2P_1}\right)\text{Ei}\left(-\frac{\lambda_1}{P_1}\right)\nonumber\\&\,\,\,\,\,\,\,\,\,\,\,\,\,\,\,\,\,\,\,\,\,\,+\frac{\lambda_1P_2}{\lambda_2P_1}e^{\frac{\lambda_2}{P_2}}\text{Ei}\left(-\frac{\lambda_2}{P_2}\right)+\frac{\lambda_1P_2}{\lambda_2P_1}-1\Bigg).
\end{align}
Here, $\text{Ei}(x)=-\int_{-x}^\infty{\frac{e^{-t}}{t}\text{d}t}$ denotes the exponential integral function. Also, $(c)$ comes from the approximation $\sqrt{1-\frac{1}{\left(1+x\right)^2}}\simeq \frac{x}{1+x}$ which is tight {for} moderate/large values of $x$, and $(d)$ is obtained by partial integration and some manipulations.
On the other hand, using (\ref{eq:eqrate2}), the instantaneous achievable rate of $\text{UE}_1$ in Slot 2 is obtained by
\begin{align}\label{eq:achiverate2}
\tilde r_1=\log\left(1+P_1G_1\right)-\frac{Q^{-1}(\theta_1)}{\sqrt{L}}\sqrt{1-\frac{1}{\left(1+P_1G_1\right)^2}}.
\end{align}
Thus, the expected rate of Slot 1, i.e., (\ref{eq:eqexpectedachiverate1}), increases to
\begin{align}\label{eq:eqexpectedachiverate2}
&E\{\tilde r_1\}\nonumber\\&=\int_0^\infty{f_{G_1}(x)\bigg(\log(1+P_1x)-\frac{Q^{-1}(\theta_1)}{\sqrt{L}}\sqrt{1-\frac{1}{\left(1+P_1x\right)^2}}\bigg)\text{d}x}
\nonumber\\&
\mathop  \simeq \limits^{(e)}  \int_0^\infty{\lambda_1e^{-\lambda_1x}\log({1+P_1x})\text{d}x}
-\frac{Q^{-1}(\theta_1)}{\sqrt{L}}\int_0^\infty{\frac{\lambda_1P_1x e^{-\lambda_1x}}{1+P_1x}\text{d}x}\nonumber\\&
=P_1e^{\frac{\lambda_1}{P_1}}\text{Ei}\left(-\frac{\lambda_1}{P_1}\right)-\frac{Q^{-1}(\theta_1)\lambda_1}{P_1\sqrt{L}}\left(e^{\frac{\lambda_1}{P_1}}\text{Ei}\left(-\frac{\lambda_1}{P_1}\right)+\frac{P_1}{\lambda_1}\right),
\end{align}
in Slot 2, where $(e)$ follows the same approximation as in (\ref{eq:eqexpectedachiverate1}).

\textbf{\underline{Note}}. As an alternative approach, not requiring the inverse $Q$ function, one can use the approximation scheme of \cite[Eq. 14]{letterzorzikhodemun}, to write 
\begin{align}\label{eq:Xapprox}
&    Q\left(\frac{\sqrt{L}\left(\log(1+u)-r\right)}{\sqrt{1-\frac{1}{(1+u)^2}}}\right)\simeq
\nonumber\\
&
\left\{\begin{matrix}
1 \text{                                          } \,\,\,\,\,\,\,\,\,\,\,\,\,\,\,\,\,\,\,\,\,\,\,\,\,\,\,\,\,\,\,\,\,\,\,\,\,\,\,\,\,\,\,\,\,\,\,\,\,\,\,\,\,\,\,\,\,\,\,\,\,\,\,\,\,\,\,\,\,\,\,\,\,\,\,\,\,\,\,\,\,\,\,\,\,\,\,\,\,\,\,\,\,\,\,\,\,\,\,\,\,\,\,\,\,\,\,\,\,\,\,\,\,\,\,\,\,\,\,\,\,\,\,\,\,\,\,\,\,\,\,\\
\,\,\,\,\,\,\,\,\,\,\,\text{if } u\le \frac{-\sqrt{\pi(e^{2r}-1)}}{2L}+e^r-1\,\,\,\,\,\,\,\,\,\,\,\,\,\,\,\,\,\,\,\,\,\,\,\,\,\,\,\,\,\,\,\,\,\,\,\,\,\,\,\,\,\,\,\,\\ 
\frac{1}{2}-\frac{\sqrt{L}}{2\pi(e^{2r}-1)}\left(u-e^r+1 \right ) \,\,\,\,\,\,\,\,\,\,\,\,\,\,\,\,\,\,\,\,\,\,\,\,\,\,\,\,\,\,\,\,\,\,\,\,\,\,\,\,\,\,\,\,\,\,\,\,\,\,\,\,\,\,\,\,\,\,\,\,\,\,\,\,\,\,\,\,\,\,\,\,\, \\\,\,\,\,\,\,\,\,\,\,\,\,\,\,\,\,\,\,\,\,\,\,\text{    if }\frac{-\sqrt{\pi(e^{2r}-1)}}{2L}+e^r-1\le u\le \frac{\sqrt{\pi(e^{2r}-1)}}{2L}+e^r-1\\ 
0  \,\,\,\,\,\,\,\,\,\,\,\,\,\,\,\,\,\,\,\,\,\,\,\,\,\,\,\,\,\,\,\,\,\,\,\,\,\,\,\,\,\,\,\,\,\,\,\,\,\,\,\,\,\,\,\,\,\,\,\,\,\,\,\,\,\,\,\,\,\,\,\,\,\,\,\,\,\,\,\,\,\,\,\,\,\,\,\,\,\,\,\,\,\,\,\,\,\,\,\,\,\,\,\,\,\,\,\,\,\,\,\,\,\,\,\,\,\,\,\,\,\,\,\,\,\,\,\,\,\,\,\,\,\,\,\,\,\,\\\text{if } u\ge \frac{\sqrt{\pi(e^{2r}-1)}}{2L}+e^r-1
\end{matrix}\right.\,\,\,\,\,\,\,\,\,\,\,\,\,\,\,\,\,\,\,\,\,\,\,\,\,\,\,\,\,\,\,\,\,\,\,\,\,\,\,\,\,\,\,\,\,\,\,\,\,\,\,\,\,\,\,\,\,\,\,\,\,\,\,\,\,\,\,\,\,\,\,\,\,\,\,\,\,\,\,\,\,\,\,\,\,\,\,\,\,\,\,\,\,\,\,\,\,\,\,\,\,\,\,\,\,\,\,\,\,\,\,\,\,\,\,\,\,\,\,\,\,\,\,\,\,\,\,\,\,\,\,\,\,\,\,\,\,\,\,\,\,\,\,\,\,\,\,\,\,\,\,\,\,\,\,\,\,\,\,\,\,\,\,\,\,\,\,\,\,\,\,\,\,\,\,\,\,\,\,\,\,\,\,\,\,\,\,\,\,\,\,\,\,\,\,\,\,\,\,\,\,\,\,\,\,\,\,\,\,\,\,
\end{align}
which simplifies the rate allocation problem of the second slot, i.e., (\ref{eq:eqrate2}), to
\begin{align}\label{eq:Xeqrate2}
      &\tilde r_1
      =\arg_x\left\{Q\left(\frac{\sqrt{L}\left(\log(1+P_1G_1)-x\right)}{\sqrt{1-\frac{1}{(1+P_1G_1)^2}}}\right)=\theta_1\right\}
      \nonumber\\
      &
      \mathop  \simeq \limits^{(f)}\arg_x\left\{   \frac{\sqrt{L}}{\sqrt{2\pi(e^{2x}-1)}}(P_1G_1-e^x+1)=\frac{1}{2}-\theta_1   \right\}
      \nonumber\\&
      =\log\left(\frac{\gamma\beta^2+\sqrt{\gamma^2\beta^4-(\beta^2-\alpha^2)(\alpha^2+\gamma^2)}}{\beta^2-\alpha^2}\right),
      \nonumber\\&
      \alpha=\frac{1}{2}-\theta_1, \gamma=P_1G_1+1,\beta=\frac{\sqrt{L}}{\sqrt{2\pi}}.
      \end{align}
Here, $(f)$ comes from (\ref{eq:Xapprox}). Also, the same procedure can be applied to find the appropriate rate allocation in Slot 1 which leads to the same result as in (\ref{eq:Xeqrate2}) except for $\gamma=\frac{P_1G_1}{1+P_2G_2}+1$. Then, one can use (\ref{eq:Xeqrate2}) and the same procedure as in (\ref{eq:eqexpectedachiverate2}) to obtain the expected system performance averaged over different channel realizations. \qedsymbol

Finally, considering the proposed approach, the following points are interesting to note:
\begin{itemize}
  \item The proposed scheme is at the cost of no extra decoding complexity at the BS, because it performs  conventional SIC-based decoding but with a different order of decoding.
  \item We presented the example for the cases with two paired UEs and two retransmissions. However, the same approach can be applied to the cases with arbitrary number of paired UEs and retransmission rounds. With different number of retransmissions, in each round the BS checks whether changing the order of message decoding gives {a} chance to improve the performance of some of the paired UEs, with no penalty for the UEs requiring retransmissions.
  \item We presented the setup for the cases with RTD HARQ. However, the same approach is applicable for different (e.g., incremental redundancy) HARQ schemes.
  \item We presented the setup for UL transmission. However, the same technique can be applied {to} the DL transmission as well. With DL transmissions, the BS informs each UE if the message of the other UE is retransmitted and the UEs adapt their message decoding orders correspondingly. This gives the BS {a} chance to increase the data rate of the UEs with successful message decoding in the previous time slot.
  \item We presented the setup for the cases where, while the message of the cell-center UE is correctly decoded, the BS fails to decode the message of the cell-edge UE. However, although it has very low probability, the same approach can be applied {to the case} with unsuccessful (resp. successful) message decoding of the cell-center (resp. cell-edge) UE as well, where the performance of the cell-edge UE is improved.
  \item We presented the setup for the cases with the same error probability constraint in successive retransmission rounds, while the same approach can be applied if the error probability constraint becomes more severe in successive retransmissions. In the cases with different error probability constraints in different retransmissions, the BS checks all possible orders of message decoding and informs the UEs correspondingly if a different order of message decoding helps to improve the system performance with no {additional} cost for the retransmitting UEs.
  \item Finally, while we presented the analysis for quasi-static conditions, the proposed scheme is also applicable in fast-fading conditions where the channel coefficients change during a (re)transmission round. This is because, even with different channel coefficients, the combination of the two received signals guarantees the error probability constraint of $\text{UE}_2$ if the initial rate is set based on (\ref{eq:eqrate1}).
\end{itemize}
\subsection{Improving the Energy Efficiency}
{In this section}, we presented the setup for the cases with rate adaption. However, the same approach can be applied {to} the cases with power adaptation in the second time slot. Particularly, considering the cases with successful (resp. unsuccessful) decoding of the message of $\text{UE}_1$ (resp. $\text{UE}_2$), in Slot 2 the BS can change the order of message decoding and, following the same discussions as in (\ref{eq:eqrate2}),  the cell-center UE can
decrease its instantaneous transmission power from $P_1$ to the one obtained by
\begin{align}\label{eq:eqpower2}
&\tilde P_1=
\arg_x\left\{\mathcal{F}\left(r_1,xG_1 \right )=\theta_1\right\}
\nonumber\\&
=\arg_x\left\{Q\left(\frac{\sqrt{L}\left(\log(1+xG_1)-r_1\right)}{\sqrt{1-\frac{1}{(1+xG_1)^2}}}\right)=\theta_1\right\}
\nonumber\\&
=\arg_x\left\{\log(1+xG_1)-r_1=\frac{Q^{-1}(\theta_1)}{\sqrt{L}}\sqrt{1-\frac{1}{(1+xG_1)^2}}\right\}
\nonumber\\&
\mathop  \simeq \limits^{(g)}
\arg_x\left\{\log(1+xG_1)-r_1=\frac{Q^{-1}(\theta_1)}{\sqrt{L}}\frac{xG_1}{1+xG_1}\right\}
\nonumber\\&
=\frac{Q^{-1}(\theta_1)+\sqrt{L}\mathcal{W}\left(-\frac{Q^{-1}(\theta_1)}{\sqrt{L}}e^{\frac{Q^{-1}(\theta_1)}{\sqrt{L}}-r_1}\right)}{\sqrt{L}{G_1}\mathcal{W}\left(-\frac{Q^{-1}(\theta_1)}{\sqrt{L}}e^{\frac{Q^{-1}(\theta_1)}{\sqrt{L}}-r_1}\right)},
\end{align}
where $\mathcal{W}(\cdot)$ is the Lambert W function and $(g)$ is obtained by the same approximation as in (\ref{eq:eqexpectedachiverate1}) which is tight at moderate/large SNRs or, equivalently, low error probability constraints, which is of our interest. 
In this way, the proposed scheme not only improves the energy efficiency of the cell-center UE but also reduces the interference for the signals of the cell-edge UE and increases its successful message decoding probability accordingly.


\section{Simulation Results}
The tightness of the approximations \cite{5452208} increases with the codeword length. For this reason, we present the results for the cases with codewords of length $\ge 100$ where, e.g., (\ref{eq:eqerrorpoly}) gives a  fairly accurate approximation of the error probability of short packets. Also, the fading parameters of the channels are set to $\lambda_1=0.1 $ and $\lambda_2=1,$ {which gives $10$ dB difference in the average received powers of the paired UEs. In the meantime, we have tested the simulations for different parameter settings and observed the same qualitative conclusions as the ones presented in the following.} Finally, by standard NOMA-HARQ we refer to the case where the order of message decoding is not updated during the retransmission rounds.

In Figs. 2-3 (resp. Figs. 4-6), we study the effect of the proposed scheme on the spectral (resp. energy) efficiency of the cell-center UE. Particularly, Fig. 2 verifies the tightness of the approximation methods of (\ref{eq:eqexpectedachiverate1}), (\ref{eq:eqexpectedachiverate2}) and (\ref{eq:Xeqrate2}), and compares the achievable rate of the cell-center UE in the cases with both the standard and the proposed NOMA-HARQ schemes. Here, the results are presented for the cases with $L=1000$ and $\theta=10^{-3}.$ Then, setting $P_1=P_2=30$ dBm, Figs. 3a and 3b demonstrate the effect of the codewords length and the UEs' error probability constraints on the achievable rate of the cell-center UE in Slot 2. 

In Fig. 4, we analyze the energy efficiency improvement of the proposed scheme in the cases with different transmission rates of the paired UEs, $L=1000, P_2=40$ dBm and based on both the simulation and analytical approach of (\ref{eq:eqpower2}). Then, Fig. 5 studies the effect of the codewords length on the required power of the cell-center UE in Slot 2 to satisfy different error probability constraints. Here, we present the result for $L=1000, P_2=30$ dBm.

\begin{figure}
\vspace{-0mm}
\centering
  \includegraphics[width=0.99\columnwidth]{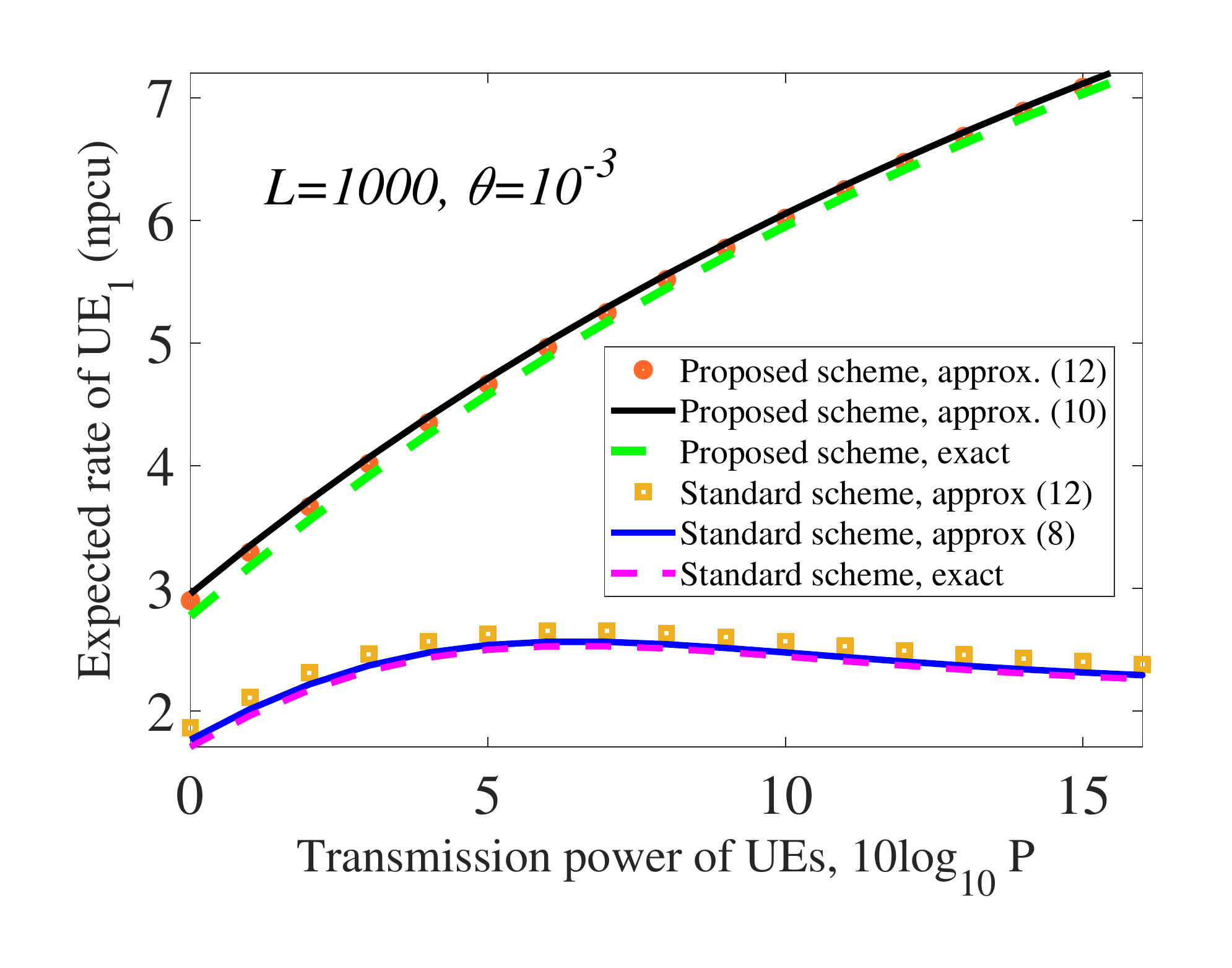}\\\vspace{-0mm}
\caption{{Achievable} rate of $\text{UE}_1$ in the second round, $L^\text{DF}=1000,$ $\theta_1=10^{-3}.$}
\vspace{-0mm}
\end{figure}

\begin{figure}
\vspace{-0mm}
\centering
 \includegraphics[width=0.99\columnwidth]{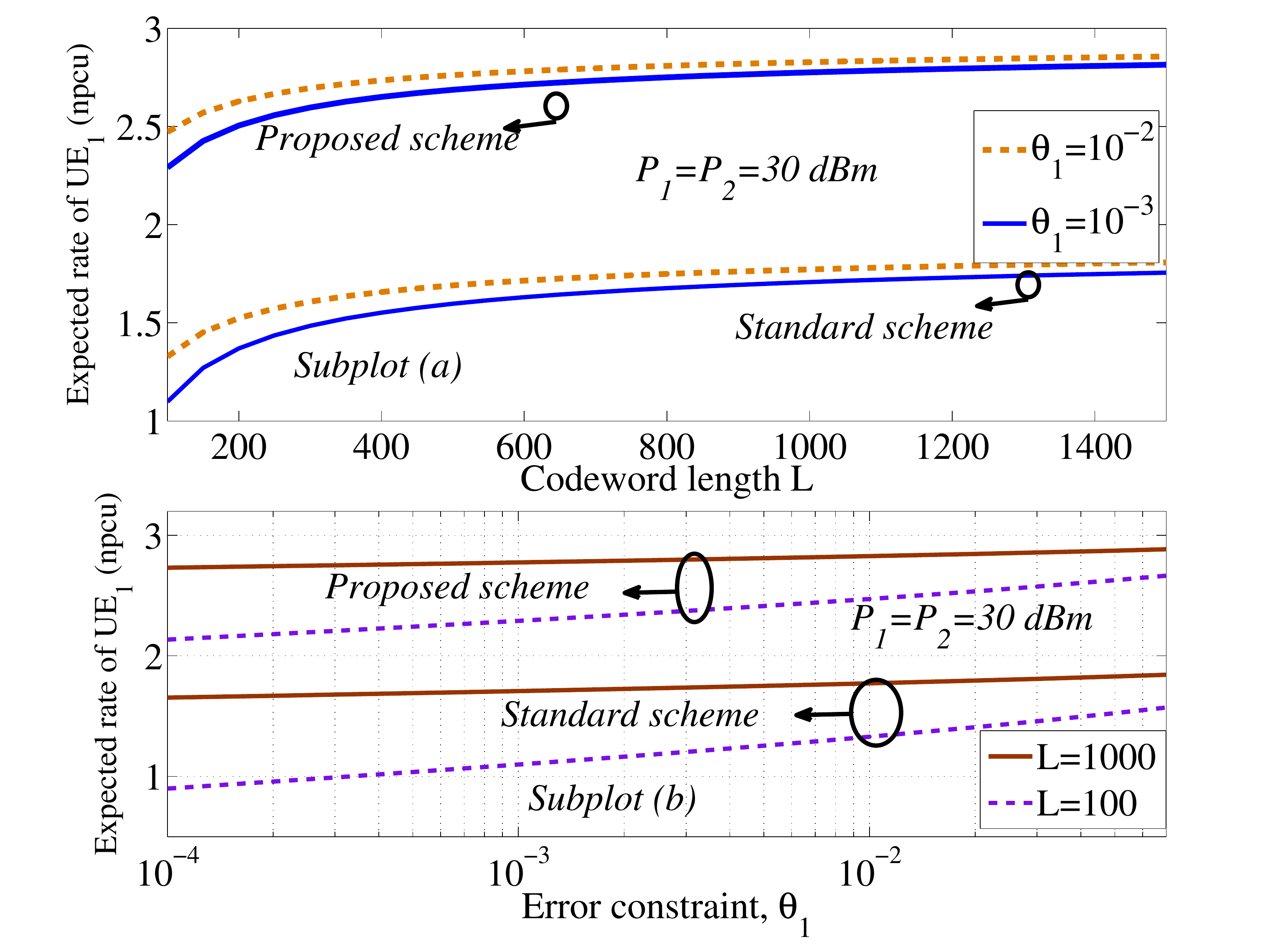}\\\vspace{-0mm}
\caption{Achievable rate of $\text{UE}_1$ in the second round versus (a) codeword length $L$ and (b) error constraint $\theta_1.$ The parameters are set as $P_1=P_2=30$ dBm.  }
\vspace{-0mm}
\end{figure}

In Figs. 2-5, we {consider} the cases where the number of information nats per codeword scales with the codeword length such that the data rate is not affected by the codeword length. On the other hand, Fig. 6 studies the expected required power of the cell-center UE in Slot 2 in the cases with a given number of information nats per codeword $K_i, i=1,2,$ and $P_2=40$ dBm, where the data rate decreases with the codewords length according to $r_i=\frac{K_i}{L},\forall i$. According to the figures, the following conclusions can be drawn:

\begin{itemize}
    \item  \emph{On the tightness of the approximations results.} As seen in Figs. 2, 4 and 5, the approximation approaches of (\ref{eq:eqexpectedachiverate1}), (\ref{eq:eqexpectedachiverate2}), (\ref{eq:Xeqrate2}) and (\ref{eq:eqpower2}) are very tight for a broad range of error constraints/parameter settings, such that the curves of the simulation and approximation results are superimposed in the figures. Thus, the proposed analytical methods can be {efficiently used} for the finite blocklength analysis of both NOMA and NOMA-HARQ protocols. 
    \item \emph{On the spectral efficiency of the proposed scheme.} With the standard error-constrained NOMA-HARQ protocol, the achievable rate of the cell-center UE is interference-limited (Fig. 2). With our proposed scheme, {instead,} the expected achievable rate of the cell-center UE increases {monotonically} with the transmission power  (Fig. 2). Finally, the efficiency of the proposed scheme, compared to standard NOMA-HARQ, increases with the UEs transmission power.
    \item \emph{On the effect of the codeword length.} Depending on whether the data rate scales with the codeword length or not, the codeword length has different effects on the system performance. First, consider the case with a given data rate, i.e., when the number of nats per codeword scales with the codeword length. Here, for both the standard and {the} proposed NOMA-HARQ protocols, the error-limited required power of the cell-center UE is slightly sensitive to the length of the codewords, if the packets are of short length. However, with a given rate, the achievable rate and the required power of the cell-center UE become insensitive to the codeword length, as the codeword length increases, and the system performance converges to {the} constant values {for} asymptotically long codewords  (Figs. 3a and 5). On the other hand, with a given number of information nats per codeword, the error-limited required power of $\text{UE}_1$ is considerably affected by the codeword length (Fig. 6). This is intuitive because, with a given number of information nats per codeword, both the error probability and the data rate decrease with the codeword length. 
    \item \emph{On the effect of the error constraint.} The sensitivity of the cell-center UE's achievable rate to the error constraint decreases with the codeword length (Fig. 3a). Moreover, compared to the standard NOMA-HARQ scheme, the proposed method is slightly less sensitive to the codeword length/error constraint (Figs. 3a and 3b). Finally, with different codeword lengths and a broad range of error constraints, the expected achievable rate of both the standard and {the} proposed NOMA-HARQ protocols scales with the error probability, in the log domain, almost linearly (Fig. 3b).
    \item \emph{On the power efficiency of the proposed scheme.} For different parameter settings, the proposed method and adapting the decoding scheme in different transmission rounds of HARQ improves the  energy efficiency of NOMA-HARQ protocols significantly. For instance, consider the parameter settings of Fig. 4, $\theta_1=10^{-3}$ and $r_1=1$ npcu. Then, compared to the standard NOMA-HARQ, our proposed scheme reduces the required SNR of the cell-center UE in the retransmission round by {$9$} dB (Fig. 4). {This is intuitive because with the proposed scheme the signal of the cell-center UE is decoded interference-free during the retransmission slots of the cell-edge UE.} Note that 1) such a performance gain is observed for a broad range of error constraints, data rates and codeword length (Fig. 4), and 2) it is at the cost of no additional receiver complexity because only the order of message decoding is updated in the retransmissions. Moreover, the relative performance gain of the proposed scheme, compared to standard NOMA-HARQ, increases with the codeword length (Fig. 6).
    {Finally, we should emphasize that 1) such spectral/power efficiency gains are achieved not in all time slots but only in the cases with retransmissions of the cell-edge UE, and 2) with a high error probability contraint, the proposed scheme may slightly increase the error probability of message decoding for the cell-edge UE. However, with moderate/low error probability constraints, which is our point of interest, the relative performance loss of the cell-edge UE is negligible (see Section III).}
    
\end{itemize}

\begin{figure}
\vspace{-0mm}
\centering
  \includegraphics[width=0.99\columnwidth]{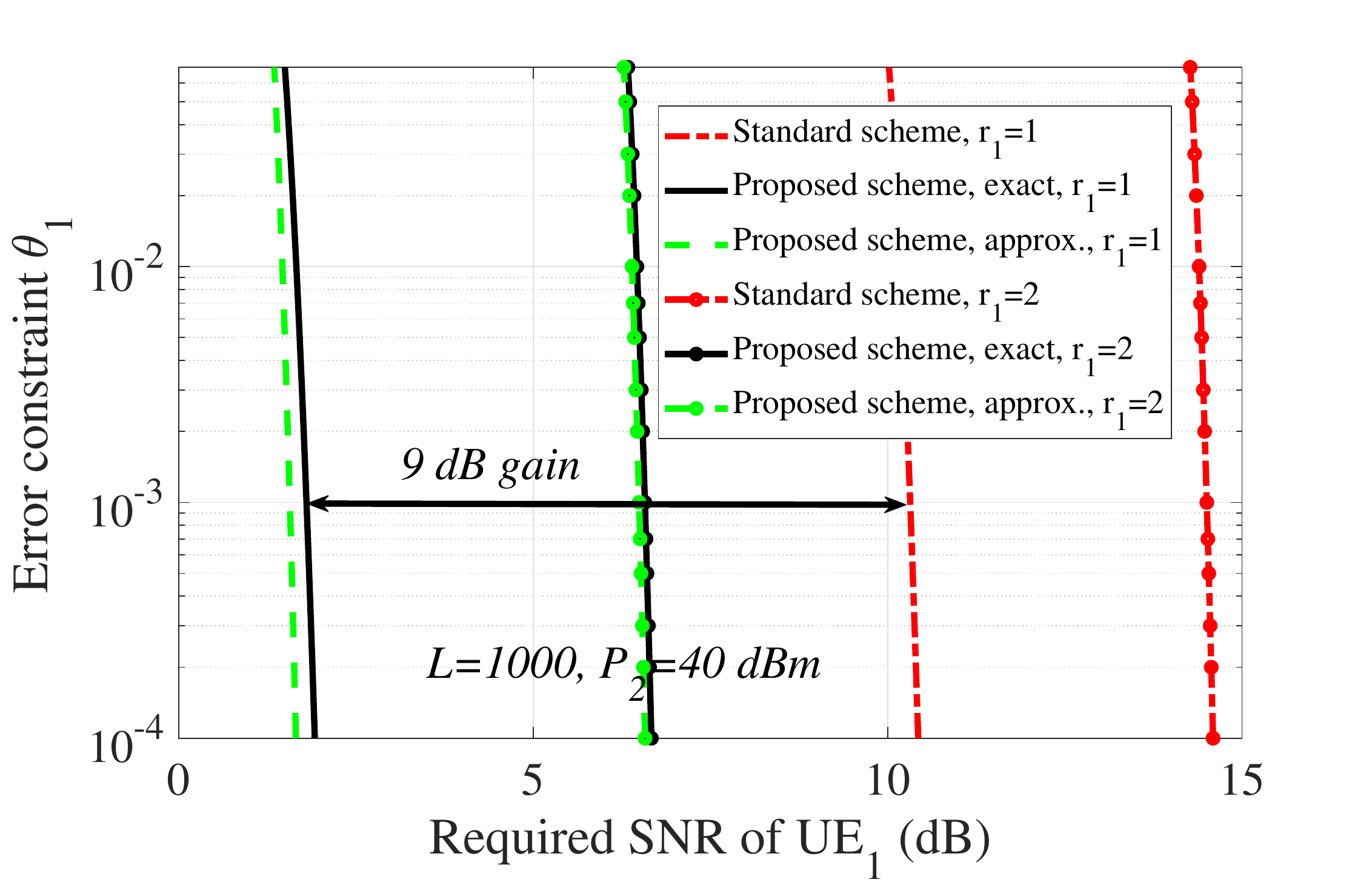}\\\vspace{-0mm}
\caption{{Power} efficiency of the proposed scheme, $L=1000,$ $P_2=40$ dBm. The results are presented for the cases with both analytical and simulation results.}
\vspace{-0mm}
\end{figure}

\begin{figure}
\vspace{-0mm}
\centering
  \includegraphics[width=0.99\columnwidth]{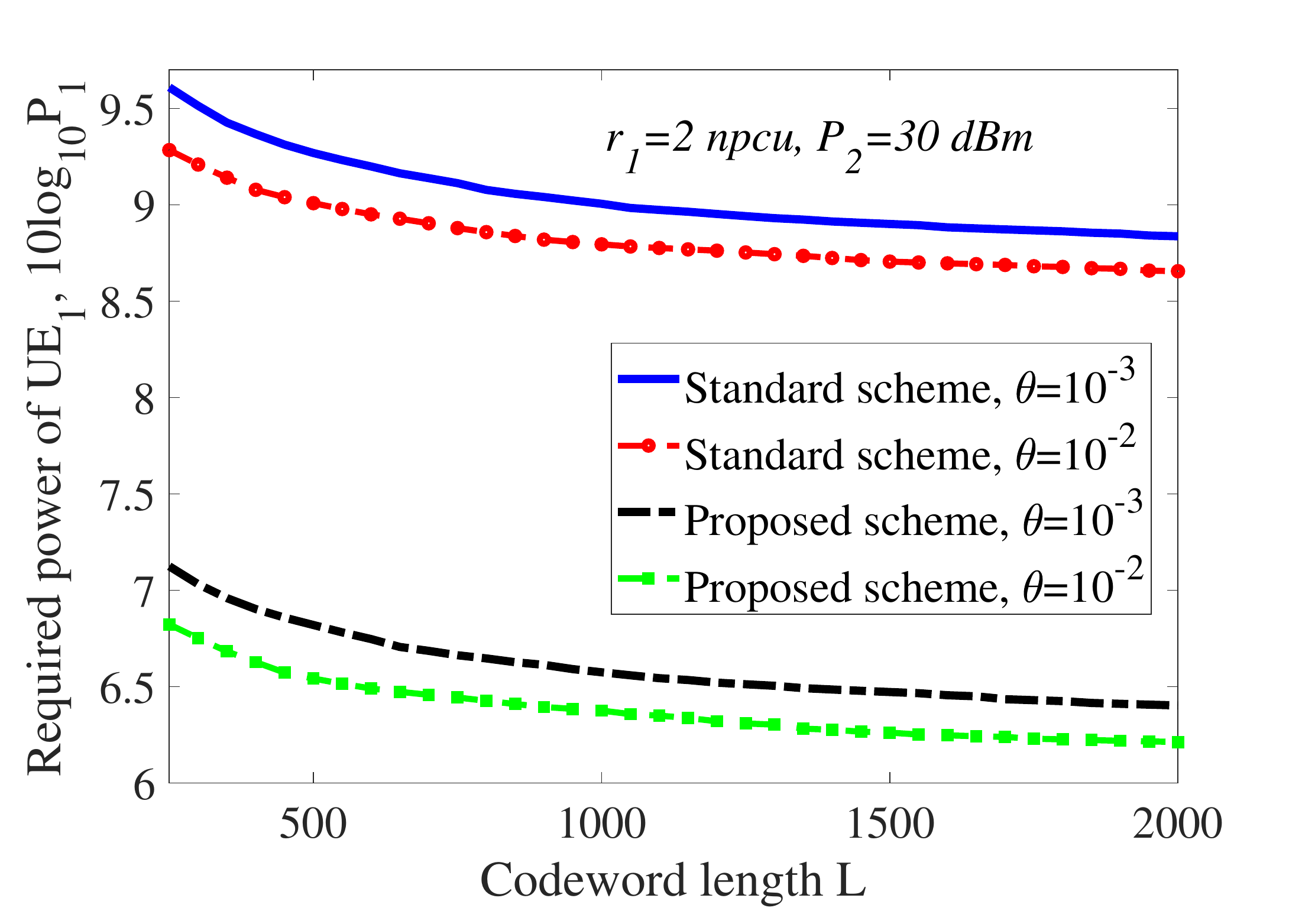}\\\vspace{-0mm}
\caption{{Required} power of the cell-center UE in the standard and proposed NOMA-HARQ schemes versus the codeword length $L$. The parameters are set as $r_i=2$ npcu, $P_2=30$ dBm. }
\vspace{-0mm}
\end{figure}

\begin{figure}
\vspace{-0mm}
\centering
  \includegraphics[width=0.99\columnwidth]{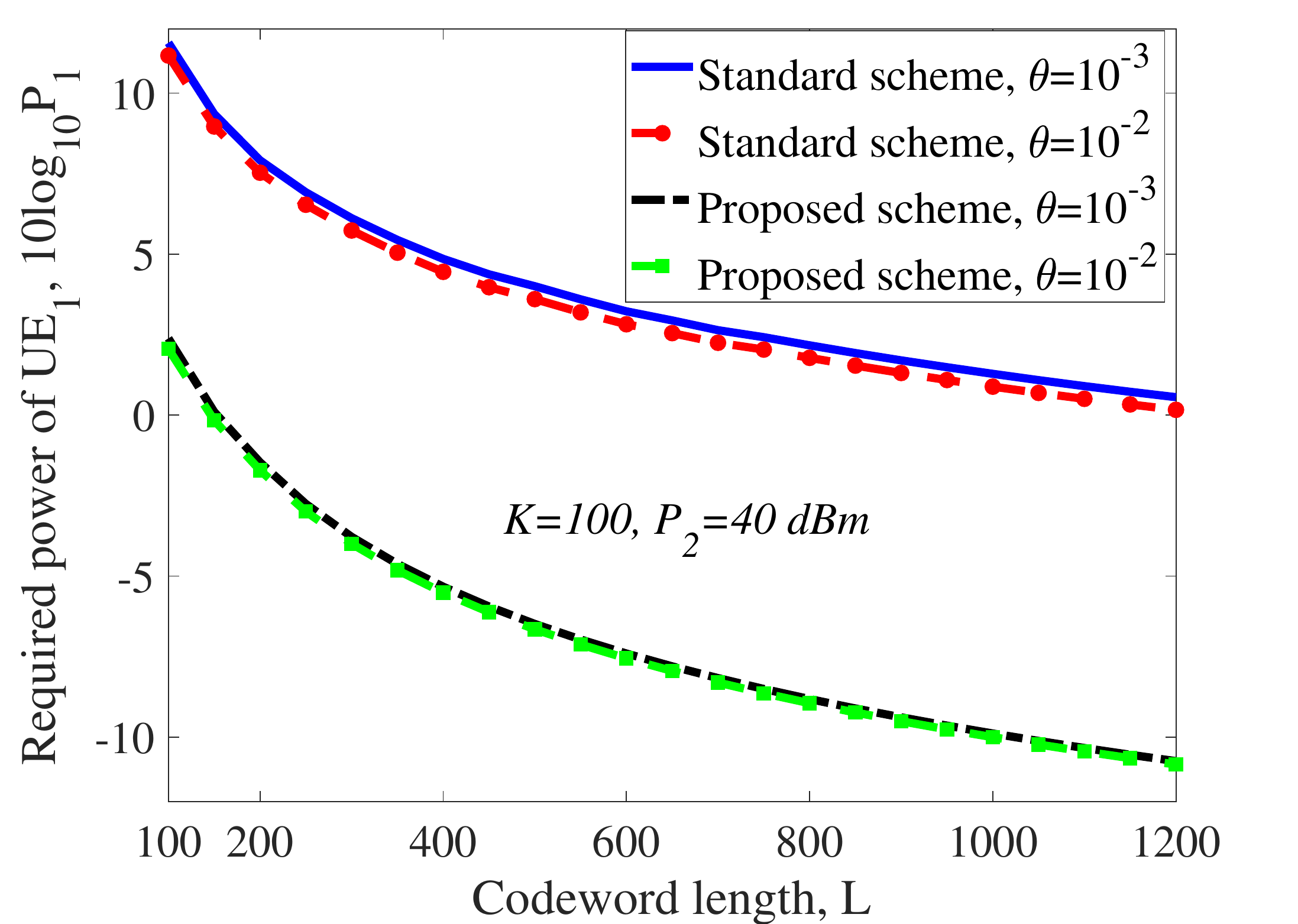}\\\vspace{-0mm}
\caption{{Required} power of the cell-center UE in the cases with a given number of nats per codeword $K$. The parameters are set as $K=100$ nats, $P_2=40$ dBm. }
\vspace{-0mm}
\end{figure}

\section{Conclusions}
We proposed and analyzed a  NOMA-HARQ approach in which adapting the decoding scheme at the receiver gives {a} chance to improve the performance of one of the paired UEs with no penalty for the other paired UE(s) and no additional implementation complexity at the receiver. The proposed scheme is fairly general in the sense that it can be applied for different HARQ protocols, both UL and DL transmission as well as for different UE pairing schemes/channel models. As we showed, with different error constraints and codeword lengths, our proposed scheme improves the spectral and energy efficiency of the cell-center UE considerably. Moreover, while the system performance is affected by the codeword length and error constraint in the cases with short packets, the effect of the codeword length decreases, as the codeword length increases.

\vspace{-2mm}
\bibliographystyle{IEEEtran} 
\bibliography{main.bib}

\begin{IEEEbiography}[{\includegraphics[width=1in,clip,keepaspectratio]{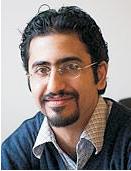}}]{Behrooz Makki} [M'19, SM'19]  received his PhD degree in Communication Engineering from Chalmers University of Technology, Gothenburg, Sweden. In 2013-2017, he was a Postdoc researcher at Chalmers University. Currently, he works as a senior researcher in Ericsson Research, Gothenburg, Sweden. 

Behrooz is the recipient of the VR Research Link grant, Sweden, 2014, the Ericsson's Research grant, Sweden, 2013, 2014 and 2015, the ICT SEED grant, Sweden, 2017, as well as the Wallenbergs research grant, Sweden, 2018. Also, Behrooz is the recipient of the IEEE best reviewer award, IEEE Transactions on Wireless Communications, 2018. Currently, he works as an Editor in IEEE Wireless Communications Letters, IEEE Communications Letters, the journal of Communications and Information Networks as well as the associate editor of Frontiers in Communications and Networks. He was a member of European Commission projects ''mm-Wave based Mobile Radio Access Network for 5G Integrated Communications'' and ''ARTIST4G'' as well as various national and international research collaborations. His current research interests include integrated access and backhaul, hybrid automatic repeat request, Green communications, millimeter wave communications, free-space optical communication, NOMA, finite block-length analysis and backhauling. He has co-authored 57 journal papers, 45 conference papers and 40 patent applications.
\end{IEEEbiography}

\begin{IEEEbiography}[{\includegraphics[width=1in,clip,keepaspectratio]{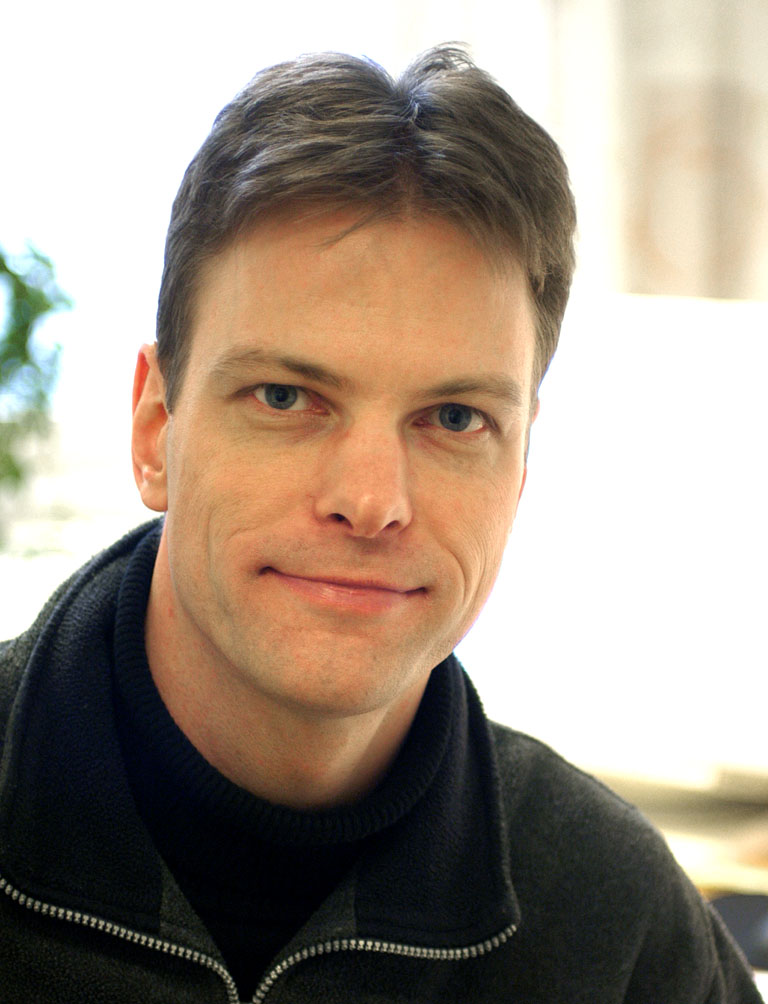}}]{Tommy Svensson} [S'98, M'03, SM'10] is Full Professor in Communication Systems at Chalmers University of Technology in Gothenburg, Sweden, where he is leading the Wireless Systems research on air interface and wireless backhaul networking technologies for future wireless systems. He received a Ph.D. in Information theory from Chalmers in 2003, and he has worked at Ericsson AB with core networks, radio access networks, and microwave transmission products. He was involved in the European WINNER and ARTIST4G projects that made important contributions to the 3GPP LTE standards, the EU FP7 METIS and the EU H2020 5GPPP mmMAGIC and 5GCar projects towards 5G and beyond, as well as in the ChaseOn antenna systems excellence center at Chalmers targeting mm-wave solutions for 5G access, backhaul/ fronthaul and V2X scenarios. His research interests include design and analysis of physical layer algorithms, multiple access, resource allocation, cooperative systems, moving networks, and satellite networks. He has co-authored 4 books, 84 journal papers, 126 conference papers and 53 public EU projects deliverables. He is Chairman of the IEEE Sweden joint Vehicular Technology/ Communications/ Information Theory Societies chapter, founding editorial board member of IEEE JSAC Series on Machine Learning in Communications and Networks, has been editor of IEEE Transactions on Wireless Communications, IEEE Wireless Communications Letters, Guest editor of several top journals, organized several tutorials and workshops at top IEEE conferences, and served as coordinator of the Communication Engineering Master's Program at Chalmers.
\end{IEEEbiography}

\begin{IEEEbiography}[{\includegraphics[width=1in,clip,keepaspectratio]{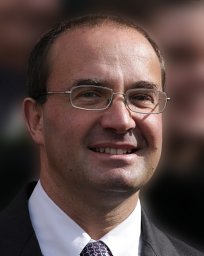}}]{Michele Zorzi} (F'07) received the Laurea and Ph.D.degrees in electrical engineering from the University of Padova in 1990 and 1994, respectively. From 1992 to 1993, he was on leave with the University of California at San Diego (UCSD). After being affiliated with the Dipartimento di Elettronica e Informazione, Politecnico di Milano, the Center for Wireless
Communications, UCSD, and the University of Ferrara, in 2003 he joined the faculty of the Information Engineering Department, University of Padova, Italy, where he is currently a Professor. His current research interests
include performance evaluation in mobile communications systems, random access in mobile radio networks, ad hoc and sensor networks, energy-constrained communications protocols, 5G millimeter-wave cellular systems, and underwater communications and networking. He was the Editor-in-Chief of IEEE Wireless Communications from 2003 to 2005, the
Editor in-Chief of the IEEE Transactions on Communications from 2008 to 2011, and the Founding Editor-in-Chief of the IEEE Transactions on Cognitive Communications and Networks from 2014 to 2018. He was a Guest Editor for several special issues of IEEE Personal Communications, IEEE Wireless Communications, IEEE Network, and IEEE Journal on Selected Areas in Communications. He served the IEEE Communications Society as a Member-at-Large in the Board of Governors from 2009 to 2011, as Director of Education and Training from 2014 to 2015 and as Director of Journals from 2020 to 2021. He received many awards from the IEEE Communications
Society, including the Best Tutorial Paper Award in 2008 and 2019, the Education Award in 2016, and the S.O. Rice Best Paper Award in 2018.
Email: zorzi@dei.unipd.it
\end{IEEEbiography}
\end{document}